\documentclass{eptcs}
\providecommand{\event}{ADG 2021} 
\RequirePackage{graphicx}
\RequirePackage{amsmath}

 \def\orcidID#1{\unskip$^{\mbox{\href{https://orcid.org/#1}{\scriptsize{[#1]}} }}$}
 
\newcommand{\abs}[1]{\lvert #1 \rvert}

\begin{document}
\title{A Method for the Automated Discovery of Angle Theorems}
\author{Philip Todd\orcidID{0000-0002-7118-8755}
\institute{Saltire Software \\ Portland, OR \\ USA}
\email{philt@saltire.com}
}

\def\titlerunning{A Method for the Automated Discovery of Angle Theorems}
\def\authorrunning{Todd}

\maketitle



\begin{abstract}
The Naive Angle Method, used by Geometry Expressions for solving problems which involve only angle constraints, represents a geometrical configuration as a sparse  linear system.  Linear systems with the same underlying matrix structure underpin a number of different geometrical theorems.  We use a graph theoretical approach to define a generalization of the matrix structure.

\end{abstract}


\section{Introduction\label{sec:Introduction}}

One approach to Geometric Discovery starts with a given geometry diagram,
and hunts systematically, or unsystematically for provable statements
about the geometric entities, or further derived geometric entities
\cite{Geogebra Discovery}. The given diagram can be in fact a parametrized family
of diagrams \cite{Polygon Discovery}. Another approach \cite{Botana Discovery} is to start with
the statements one wants to prove, and discover supplementary conditions
required to make the theorems true. Again, however, the geometric
milieu is given.  A problem for such systems is to determine the interestingness of
generated theorems, metrics for which are an active topic of research\cite{Interestingness}.

In this paper, we consider working in reverse and generating
the geometric diagram to match a more abstract form of the
theorem, which guarantees both its solution, but also a certain level of interestingness. The abstract form is developed by analogy with known theorems,
considered (by this author) to be aesthetically pleasing. We develop
and automate here a method for generating many theorems of comparable
structure but different geometry to our seed theorems. Hopefully this
might lend us some control of the richness and tractability, even
aesthetic appeal of our generated theorems.

Having promised emergent geometry, we immediately limit the scope
of our work, however, to consider theorems in the Naive Angle Method
employed by Geometry Expressions \cite{GX} for angle specific problems. While
the method accommodates a number of different constraint types, in
the bulk of this paper, we focus solely on the angle bisector constraint,
which can be disguised as an isosceles triangle, a circle chord,
or a reflection. In any case, it contributes a row with 3 values -1,-1,2
to the constraint matrix. At one level, we can re-interpret the same
matrix using different geometry: for example changing a circle chord
into an angle bisector (figure \ref{fig:ZhangTheorem}). At another level, we consider matrices
with non zero elements in the same places, but with different assignments
within the row of the numerical values (they will still be -1, -1
, 2, only their order will be different). At a third level, we generalize to consider matrices with a similar pattern
of non-zero positions. For a class of such matrices, we give structural
conditions which determine the presence or absence of theorems of
comparable interest to the prototype.

\section{The Naive Angle Method}

The Full Angle Method \cite{Full Angle} treats an angle and its supplement as the same
thing. It has the benefit of allowing theorems to be expressed in
a more general form, and hence allowing multiple instances of the
same theorem to be proved at once \cite{Zhang}. However, it does not
allow theorems to be expressed which themselves depend on the conventional
notion of an angle, as the difference in the direction of two signed
rays. Geometry Expressions \cite{GX} employs the Analytical Geometry
Method \cite{GXAG} to derive an expression for output measurements in terms of
symbolic values for input constraints. The method relies
on deriving a symbolic Cartesian coordinate description of the model.
When an angle is measured, it is derived from the Cartesian equations
of its lines using inverse trigonometric functions. Simplification
of expressions involving such functions imposes a heavy burden on
the algebra system, and results for angle-dominated diagrams are not
satisfactory. This prompted the development of an auxiliary system
which is deployed when the entire computation can be kept in the angle
domain. For example, if a triangle is defined by one side and two
angles, and the third angle is measured, the Cartesian computation
involves an inverse trigonometric function. However, expressed in terms of angles the result
is simply the difference between $\pi$ and the sum of the other two angles.
In contrast to the full angle method, this approach considers angles
to be signed, and hence an angle and its supplement are different.
This means that theorems are less general, but has the advantage of
corresponding with the conventional notion of angle used by consumers
of mathematics such as students and engineers. 

Let $d_{1}\ldots d_{n}$ be the directions of the $n$ (directed)
lines comprising a geometric figure. A number of different constraints
may be applied in Geometry Expressions, each of which may be expressed
as a linear equation: 
\begin{enumerate}
\item angle between line $i$ and $j$ is $\phi$ : $d_{i}-d_{j}=\phi$
\item line $k$ bisects line $i$ and line $j$ : $2d_{k}-d_{i}-d_{j}=0$
\item line $k$ is the base of an isosceles triangle whose equal sides are
$i$ and $j$ : $2d_{k}-d_{i}-d_{j}=\frac{\pi}{2}$
\item line $j$ is the image of line $i$ under reflection in $j$ : $2d_{k}-d_{i}-d_{j}=0$
\end{enumerate}
Circles contribute to the angle model in two ways:
\begin{enumerate}
\item Tangents contribute right angles with the line joining the point of
tangency to the center of the circle.
\item Chords contribute isosceles triangle relationships with their end
point radii.
\end{enumerate}
The linear equations above are all to be considered modulo $2\pi$.
When the resulting linear system is solved, symbolic directions are
determined for each line, which can be subtracted to yield angles.
Resolving how many $2\pi$'s to add or subtract from the result is
done by preserving a numerical prototype of the geometry (a sketch)
which is used to arbitrate this issue.

To solve for the value of an angle, the columns of the matrix are reordered
so that the columns corresponding to the two lines which define the
angle are at the right. Gaussian Elimination is then run forward to
create an upper diagonal matrix. If the Gaussian Elimination runs
to completion with non-zero values in the last two places of the final row and zeros in
the rest then the angle is determined. 

Any row constructed by Gaussian Elimination is in the null space of
the row vectors representing the constraints on the problem. Establishing
that a row representing a specific angle is in that null space is
equivalent to proving that the value of the angle may
be determined from the constraint equation. The actual value of the
angle will be generated in the course of the Gaussian Elimination
applied to an additional matrix column containing the right hand sides
of the constraint equations.

\section{Theorem Discovery}

A statement in the Naive Angle Method is a linear combination of line directions. 
It can be proven true and constitutes a theorem if it is in the row space of the hypothesis vectors.
If there are $m$ rows (hypotheses) and $n$ columns (geometric
lines), then, assuming the hypotheses are linearly independent, they
span a space of dimension $m$ from a space of dimension $n-1$ (all rows
satisfy the linear condition that the sum of their coefficients is
0). Let $C$ be a set of $n-m+1$ or more columns of the matrix (or equivalently,
geometric lines). Then there is a row vector in the span of the hypothesis
vectors whose non-zero coefficients are all in $C$. 

For any set $C$ of sufficient size, then, a theorem exists. Some are more interesting than others. For a given diagram, we are most interested
in theorems which use all the hypotheses: that is their vector does
not belong to a space spanned by any subset of the hypothesis vectors.
A theorem vector is less common, and thus more interesting, the fewer
non-zero coefficients it contains. An algorithm for finding interesting
theorems in a given diagrams takes the following form:
\begin{enumerate}
\item Augment the matrix $M$ with a final column containing the symbols $r_{1},\ldots r_{m}$ 
\item Iterate through a space of potential column sets $C$ , either exhaustively,
or using random permutations.
\item Reorder the matrix columns so that the final $n-m+1$ columns correspond
to the indices in $C$.
\item Perform Gaussian Elimination to compute the upper triangular matrix
U in the LU decomposition.
\item The final row of U contains the theorem in its first n entries and
its expression as a linear combination of rows in the final column.
\item decide which 'theorems' to keep based on a heuristic which can use
the number of hypotheses involved and the number of non zero coefficients
in the theorem.
\end{enumerate}

\subsection{Matrix Structure of Seed Theorems }

The approach above starts with a geometry diagram which then defined
a matrix, we took the matrix and used it to generate quantities which
were guaranteed to be true, then applied heuristics to determine which
of these true facts was worth holding onto as a theorem. We'd like
to consider starting with a matrix and generating the geometry diagram
from the matrix. As there are three different constraints which contribute
a row with values -1, -1, 2, such a matrix row can be interpreted
in those three different ways. Further, isosceles triangle constraints
may be represented by a line forming the chord of a circle.

\begin{figure}
\begin{center}
\includegraphics[scale=0.35]{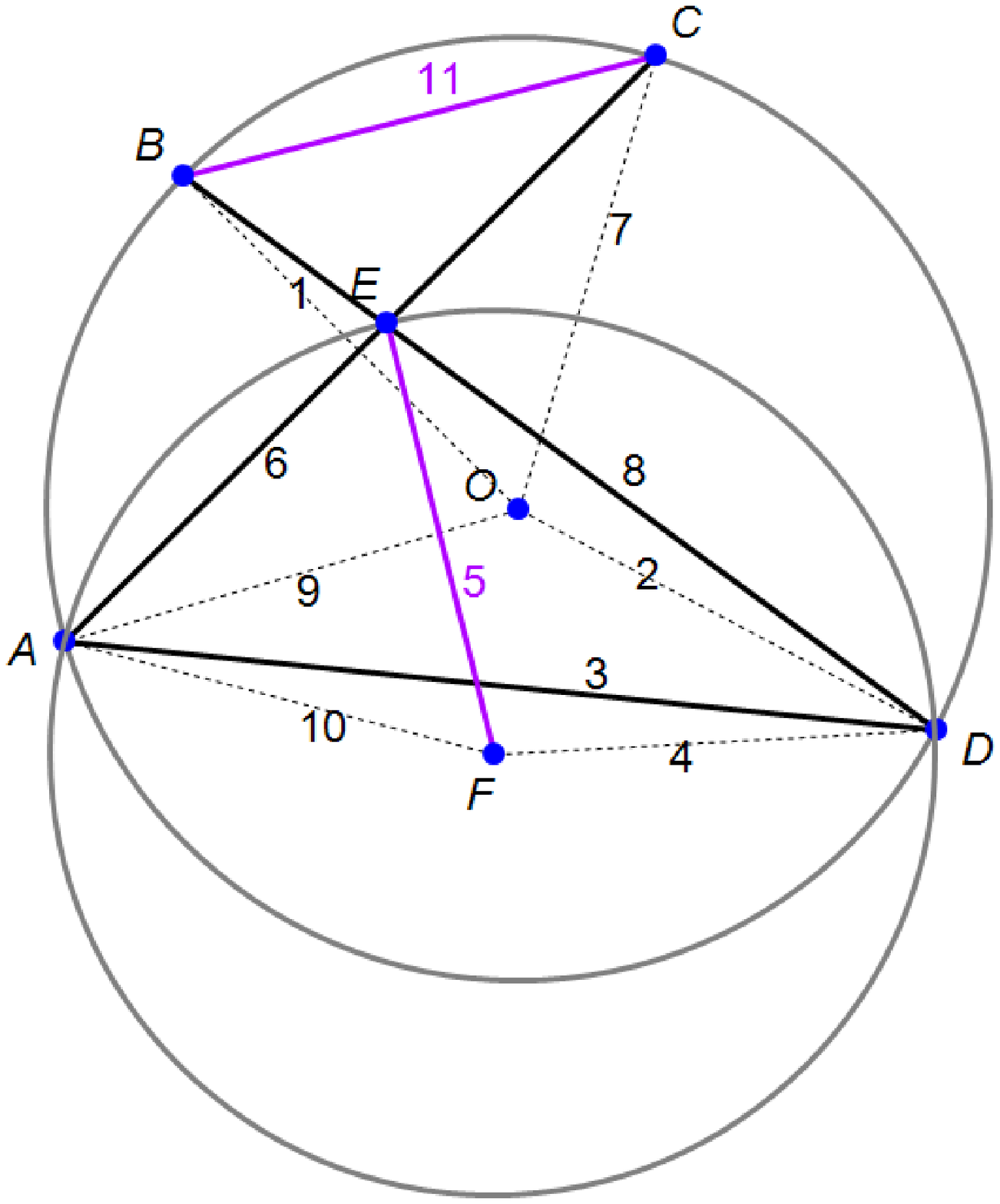} \includegraphics[scale=0.35]{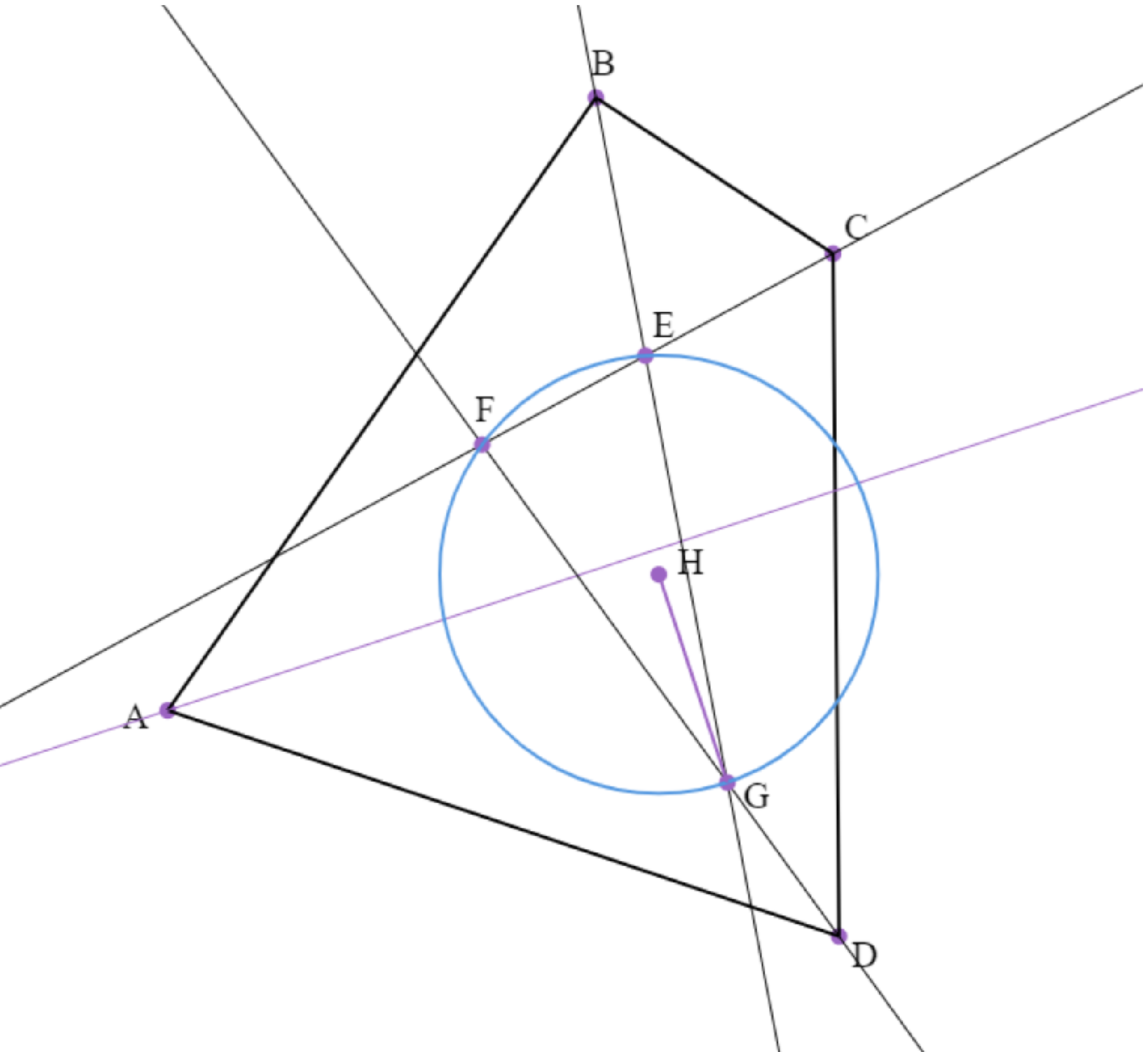}
\end{center}
\caption{\label{fig:ZhangTheorem}Two theorems which share the same matrix}

\end{figure}

Hence the same matrix can represent two apparently quite different
diagrams. For example, in figure \ref{fig:ZhangTheorem}(a) $E$ is the intersection
of the diagonals of cyclic quadrilateral $ABCD$, $F$ is the circumcenter
of $ADE$, the theorem states that $BC$ and $EF$ are perpendicular \cite{Zhang}. Radial
lines are added as needed in the Naive Angle Method and are shown
dashed in the diagram.

Numbers on the diagram reference matrix columns. The corresponding
matrix is shown below

\[
\left[\begin{array}{ccccccccccc}
-1 & 0 & 0 & 0 & 0 & 0 & -1 & 0 & 0 & 0 & 2\\
-1 & -1 & 0 & 0 & 0 & 0 & 0 & 2 & 0 & 0 & 0\\
0 & -1 & 2 & 0 & 0 & 0 & 0 & 0 & -1 & 0 & 0\\
0 & 0 & 2 & -1 & 0 & 0 & 0 & 0 & 0 & -1 & 0\\
0 & 0 & 0 & -1 & -1 & 0 & 0 & 2 & 0 & 0 & 0\\
0 & 0 & 0 & 0 & -1 & 2 & 0 & 0 & 0 & -1 & 0\\
0 & 0 & 0 & 0 & 0 & 2 & -1 & 0 & -1 & 0 & 0
\end{array}\right]
\]

Figure \ref{fig:ZhangTheorem}b shows a second theorem which has the same matrix. $ABCD$ is
a general quadrilateral. The angle bisector at $C$ intersects the angle
bisectors at $B$ and $D$ in $E$ and $F$, while the angle bisector at $B$ intersects
the angle bisector at $D$ in $G$. $H$ is the circumcenter of $EFG$. The theorem
states that $HG$ is perpendicular to the bisector of angle $A$. Rows 1,2,3,6
are interpreted in figure \ref{fig:ZhangTheorem}b as angle bisectors, while in figure \ref{fig:ZhangTheorem}a they
are considered to be isosceles triangle constraints.

\subsection{Different Matrices with the Same Shape}

\begin{figure}
\begin{center}
\includegraphics[scale=0.35]{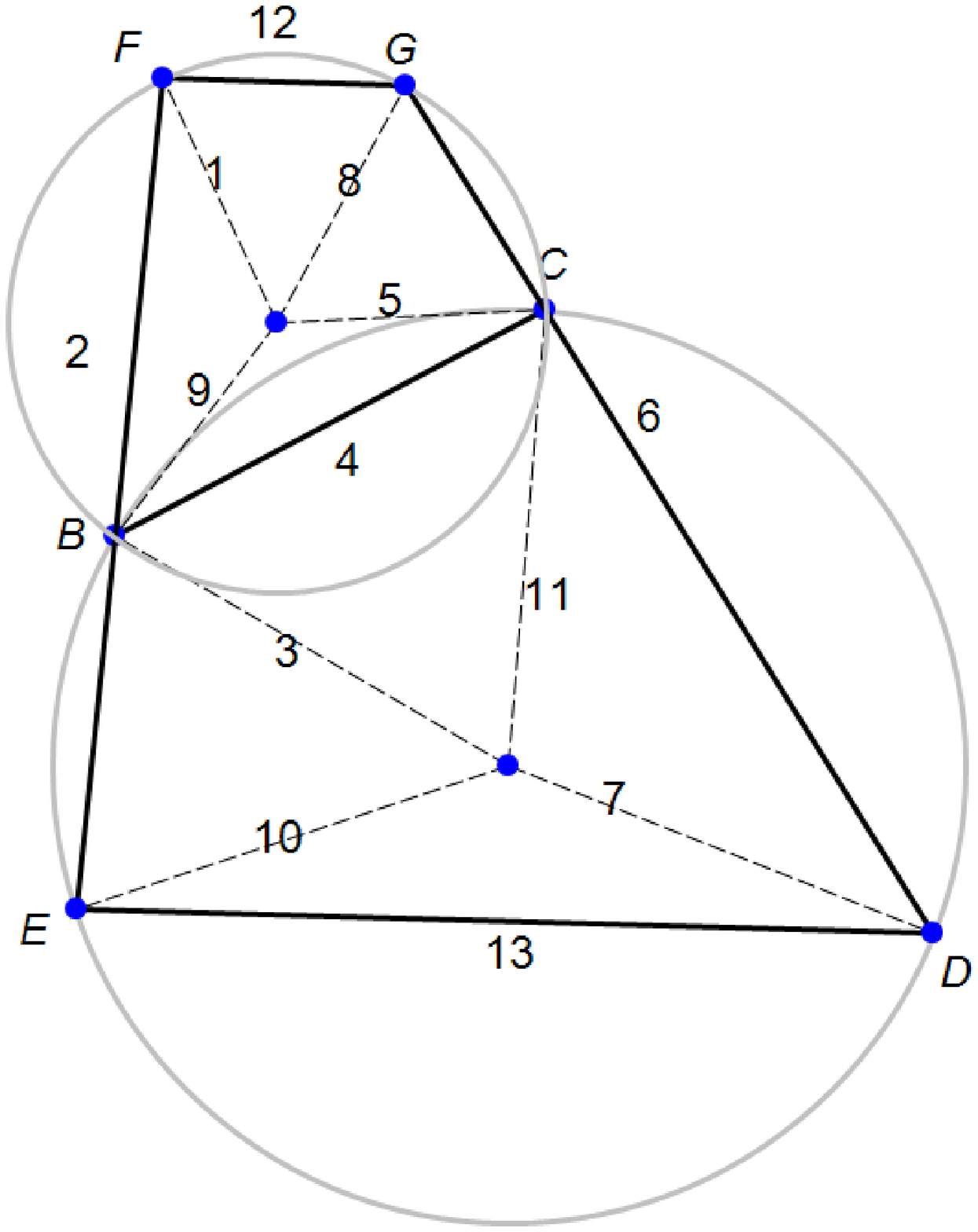}
\includegraphics[scale=0.35]{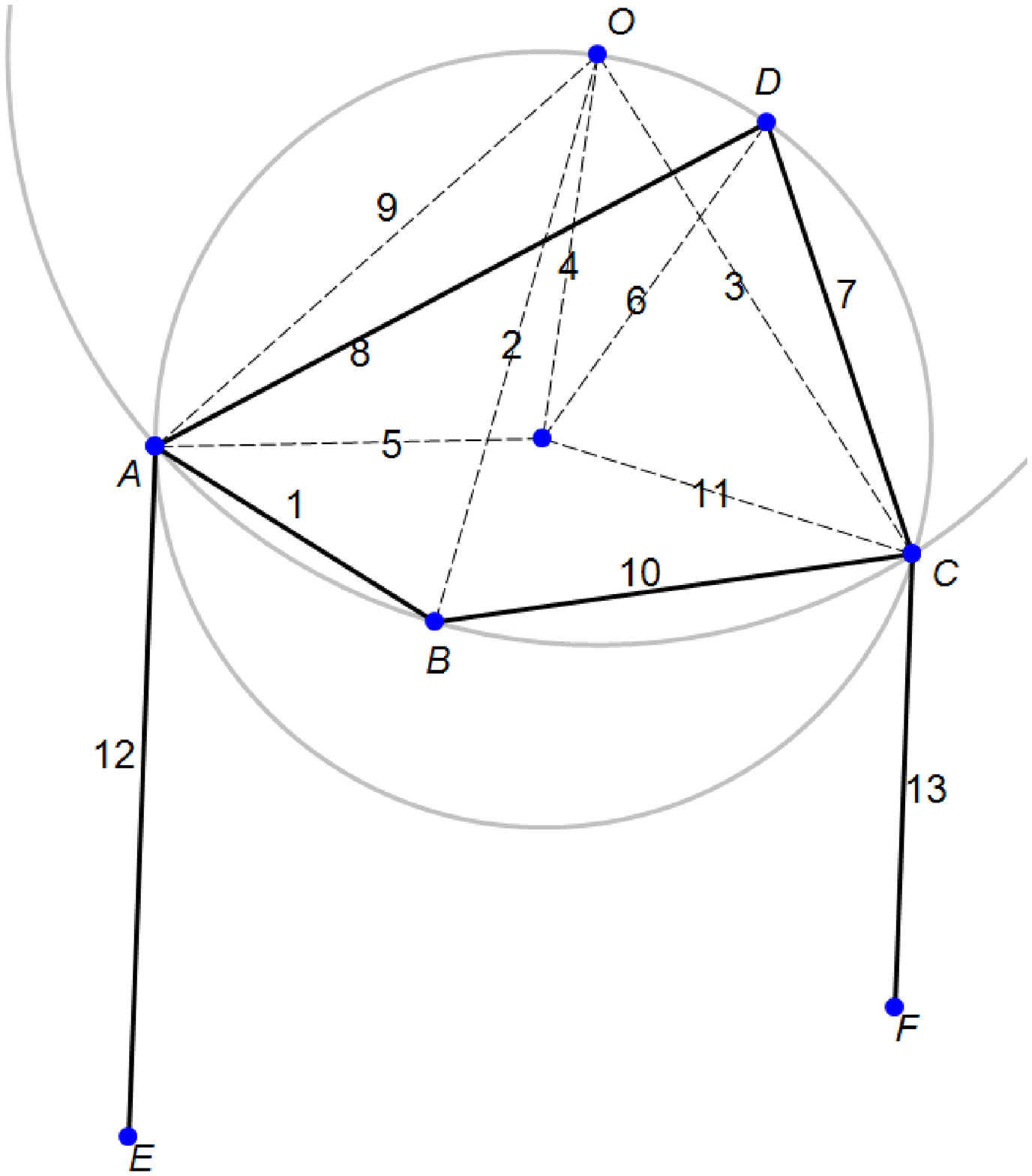}
\end{center}

\caption{\label{fig:TwoCyclicQuads} Two theorems whose matrices have non-zero elements in the same locations,
but which are not identical}

\end{figure}

Figure \ref{fig:TwoCyclicQuads}a  states that if $BCDE$ is a cyclic quadrilateral
and $F$ lies on $EB$ extended while $G$ lies on $DC$ extended, and $BFGC$ is
also cyclic, then $FG$ and $ED$ are parallel \cite{Zhang}. Radial lines, added automatically
in the Naive Angle Method, are shown dashed. Numbers in the diagram
correspond to columns in the matrix.

\[
\left[\begin{array}{ccccccccccccc}
-1 & 0 & 0 & 0 & 0 & 0 & 0 & -1 & 0 & 0 & 0 & 2 & 0\\
-1 & 2 & 0 & 0 & 0 & 0 & 0 & 0 & -1 & 0 & 0 & 0 & 0\\
0 & 2 & -1 & 0 & 0 & 0 & 0 & 0 & 0 & -1 & 0 & 0 & 0\\
0 & 0 & -1 & 2 & 0 & 0 & 0 & 0 & 0 & 0 & -1 & 0 & 0\\
0 & 0 & 0 & 2 & -1 & 0 & 0 & 0 & -1 & 0 & 0 & 0 & 0\\
0 & 0 & 0 & 0 & -1 & 2 & 0 & -1 & 0 & 0 & 0 & 0 & 0\\
0 & 0 & 0 & 0 & 0 & 2 & -1 & 0 & 0 & 0 & -1 & 0 & 0\\
0 & 0 & 0 & 0 & 0 & 0 & -1 & 0 & 0 & -1 & 0 & 0 & 2
\end{array}\right]
\]

Figure \ref{fig:TwoCyclicQuads}b illustrates the following theorem from geometrical optics.  
The image of parallel rays reflecting at points $A$ and $C$ in the sides $AB$ and $BC$ of triangle $ABC$
lies on the circumcircle of $A$, $B$ and $O$, where $O$ is the circumcenter of $ABC$.
With the columns numbered as
given in the figure, the matrix for this diagram is as follows:

\[
\left[\begin{array}{ccccccccccccc}
2 & 0 & 0 & 0 & 0 & 0 & 0 & -1 & 0 & 0 & 0 & -1 & 0\\
2 & -1 & 0 & 0 & 0 & 0 & 0 & 0 & -1 & 0 & 0 & 0 & 0\\
0 & -1 & -1 & 0 & 0 & 0 & 0 & 0 & 0 & 2 & 0 & 0 & 0\\
0 & 0 & 2 & -1 & 0 & 0 & 0 & 0 & 0 & 0 & -1 & 0 & 0\\
0 & 0 & 0 & -1 & -1 & 0 & 0 & 0 & 2 & 0 & 0 & 0 & 0\\
0 & 0 & 0 & 0 & -1 & -1 & 0 & 2 & 0 & 0 & 0 & 0 & 0\\
0 & 0 & 0 & 0 & 0 & -1 & 2 & 0 & 0 & 0 & -1 & 0 & 0\\
0 & 0 & 0 & 0 & 0 & 0 & -1 & 0 & 0 & 2 & 0 & 0 & -1
\end{array}\right]
\]

Examining these matrices, we see that both have non-zero elements
in the same positions of the matrix, however the values in those elements
are not identical. The matrices have the same shape, but are not identical.

As the final two columns in this matrix have only one entry, any theorem
which uses all hypotheses must incorporate these two columns. At this
point, we narrow our focus to theorems whose statement as a linear
combination of line directions contain only two non-zero coefficients.
As the sum of each row of our matrix is 0, then a theorem with 2 non
zero coefficients must specify a scalar multiple of the difference
between the line directions, or the angle between the lines.

With this narrower focus, matrices of this shape which yield theorems
can be determined as follows:
\begin{enumerate}
\item Iterate through all $3^{8}$ different locations for the 2 in each
row.
\item Perform Gaussian Elimination to determine the upper triangular component
of the LU decomposition.
\item Examine its last row: if it has 0's in columns 7-11 our theorem exists.
Otherwise it does not.
\end{enumerate}
Running this algorithm on the matrix shape shown yields 33 which define
theorems determining the angle between lines 12 and 13. 

\section{Generalized Theorem Structure}

\begin{figure}
\begin{center}
\includegraphics[scale=0.6]{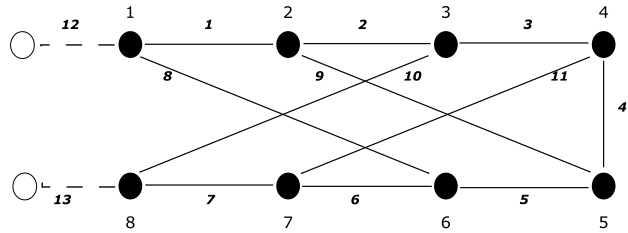}
\end{center}

\caption{\label{fig:Graph8}Graph representation of the matrix structure for Figure 2 }
\end{figure}

We have taken a matrix which is known to generate a worthwhile theorem,
and interpreted it in a different way geometrically. We have further
taken a matrix whose shape is shared by two interesting theorems, varied the numeric values attributed to the
non-zero entries and generated a collection of theorems sharing the matrix shape. 
We now would like to construct other matrices which
have a good chance of generating interesting theorems. We narrow our
consideration to matrices formed solely by the constraints 2, 3 and
4 above. Hence each hypothesis contributes a row to the matrix formulation
of the problem comprising three entries (two -1's and a 2). The matrices
derived from the theorems of figure \ref{fig:TwoCyclicQuads} also have the following characteristics:
\begin{itemize}
\item All but two columns of the matrix contain 2 non-zero elements.
\item Two rows both have non-zero entries in at most a single column.
\end{itemize}
Such a theorem can be represented as a graph in the following way
(figure \ref{fig:Graph8}). Each vertex of the graph represents a row of the matrix.
An edge joins two vertices if there are two non-zero entries in a
column. If a column contains a single non-zero entry, then a node
is added and an edge joining it to the node representing the row of
the non-zero entry. In the figure, the numbers on the graph vertex
correspond to row numbers in a matrix, whereas the numbers on the
graph edges correspond to column numbers. This graph represents the
matrix for the diagrams of figure \ref{fig:TwoCyclicQuads}. The two dashed lines attached
to the 1st and 8th nodes correspond to the matrix columns with single
non-zero entries, and thus to geometric lines whose angle would be
determined. These nodes (unfilled in the diagram) do not correspond
to matrix rows.

An associated cubic graph (uniform degree 3) may be defined by removing the white nodes
and dashed edges and adding an edge between the two vertices with
degree 2. Equivalently, in the matrix representation, we merge the
two columns which have a single non-zero entry. 

Reversing this process, given a cubic hamiltonian graph $H$, we create
a graph $G$ by removing one edge. Let $u$ and $v$ be the graph nodes
adjacent to this edge. These have degree 2 in $G$. We number the nodes
of $G$ from $1$ to $m=2p$ and its edges from $1$ to $n=3p-1$. We
construct a matrix $P$ such that for $j\leq n$ $P_{ij}=1$ if vertex
$i$ is adjacent to edge $j$ and $0$ otherwise. We add two colums
defined as follows:

\[
P_{i,n+1}=\begin{cases}
1 & i=u\\
0 & otherwise
\end{cases}
\]

\[
P_{i,n+2}=\begin{cases}
1 & i=v\\
0 & otherwise
\end{cases}
\]

A potential theorem matrix $M$ is formed by assigning values to the non zero entries of the pattern matrix
$P$. Given such a pattern with $m$ rows, there are clearly $3^{m}$
different such assignments. 

When converting
a matrix to geometry, there is a choice for each row of the matrix
whether to interpret the row as an angle bisector (equivalently a
reflection) or as an isosceles triangle. If two isosceles triangles
share one of the equal sides, this can be represented geometrically
as two chords of a circle meeting at a point. The line from the circle's
center to the common point may be omitted from the diagram, and is
implied by the geometry. If there are several opportunities to use
the same circle, this can make the diagram and hence the theorem,
much more appealing. In particular, we form the graph $G'$
whose vertices correspond to the columns of the matrix, and whose
edges correspond to the rows: each edge joins the two columns in that row whose
values are -1. Cycles in $G'$ can be made to correspond to cyclic
polygons in the geometry figure and much clutter disappears.

\begin{figure}
\begin{center}
\includegraphics[scale=0.4]{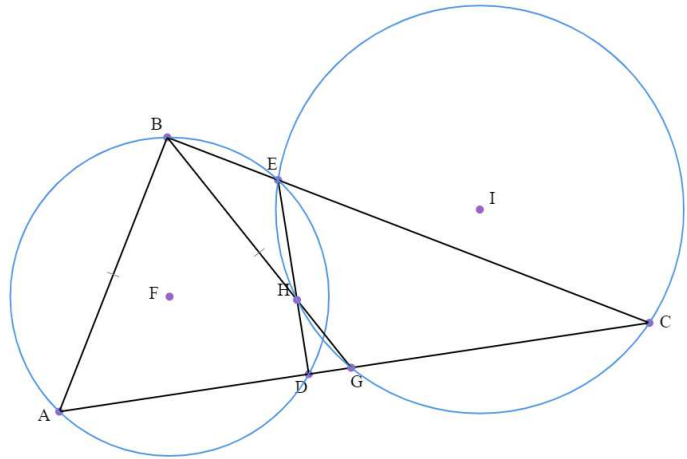}
\end{center}
\caption{\label{fig:Theorem10}A Theorem Corresponding to a generated matrix. $ABC$ is a triangle $E$
and $D$ lie on sides $BC$ and $AC$ such that $ABED$ is cyclic. $G$ is on $AC$
such that $\abs{ AB} =\abs{BG}$. 
$H$ is the intersection of $BG$ and $DE$. Then $C,E,H,G$ are cyclic.}

\end{figure}

As an example, a theorem generated from the 10 node pattern is depicted
in figure \ref{fig:Theorem10}.

\section{Conclusion}

Our approach to theorem discovery is to identify matrix patterns which,
with appropriate numeric values lead to theorems. In this paper, we
have narrowed our definition of what constitutes a 'theorem' to be
that an angle between a pair of lines is determined by the hypotheses.
We have also narrowed our focus to theorems which are described solely
in terms of angle bisector (equivalently isosceles triangle) constraints.

The rows of our matrices, corresponding to bisector constraints, contain
three non-zero entries, and their values are 2, -1, -1. We further
specialize by specifying that all but one or two columns must contain exactly
two non zero elements, and that column or columns should contain one.
The locations of the non zero elements in matrices of this kind can
be represented as a graph where each row corresponds to a graph node,
and each column with two non-zero elements corresponds to a graph
edge between the corresponding nodes. 

A straightforward mechanistic approach to converting the matrix to a geometry theorem
would be to consider every constraint to be simply an angle bisector.
With this approach, the matrix of figure \ref{fig:ZhangTheorem} yields the following theorem.

Given 11 lines A, B, C, D, E, F, G, H, I, J, K such that K is the angle bisector of A and G, 
H is the angle bisector of A and B and of D and E, C is the angle bisector of B and I and of D and J,
and F is the angle bisector of E and J and of G and I, then lines E and K are parallel.

While straightforward to mechanize, the resulting theorem is far from elegant.  Judicious choice of
geometric representation for each matrix row, between isosceles triangle constraint, 
angle bisector, circle chord and reflection can greatly enhance the attractiveness of the theorem, as can 
considerations of symmetry.  The development of a set of heuristics
to drive the automation of this choice  is a topic for further work.

A potential use of this capability, once developed would be for the
automated generation of non-trivial proof problems for students. If
a problem comes from a specific graph shape, its level of complexity
could be controlled.

\providecommand{\urlalt}[2]{\href{#1}{#2}}
\providecommand{\doi}[1]{doi:\urlalt{http://dx.doi.org/#1}{#1}}

\bibliographystyle{eptcs}

\end{document}